\documentclass[aps,twocolumn]{article}

\usepackage{amssymb}
\begin{document}
\author{Li-Xiang Cen$^1$\thanks{%
Electronic address: gontp@spin.lzu.edu.cn} and Shun-Jin Wang$^{1,2}$ \\
%EndAName
{\small 1}{\it . }{\small Department of Modern Physics, Lanzhou University,
Lanzhou 730000, P.R.China}\\
{\small 2}. {\small Institute of Modern Physics, Southwest Jiaotong
University, Chengdu 610031, P.R.China}}
\title{Distilling a Greenberger-Horne-Zeilinger state from an arbitrary pure
state of three qubits}
\date{}
\maketitle

\begin{abstract}
We present a general algorithm to achieve local operators which can produce
the GHZ state for an arbitrary given three-qubit state. Thus the
distillation process of the state can be realized optimally. The algorithm
is shown to be sufficient for the three-qubit state on account of the fact
that any state for which this distillation algorithm is invalid cannot be
distilled to the GHZ state by any local actions. Moreover, an analytical
result of distillation operations is achieved for the general state of three
qubits.
\end{abstract}

%\date{}

\vskip0.2in \noindent PACS numbers: 03.67.-a, 03.65.Bz, 03.67.Hk \vskip0.2in

Entanglement manipulation is an important issue in the studies of quantum
information theory. On the one hand, it is related to the basic problem of
which tasks one can accomplish with a given resource of entanglement \cite
{bennett1}-\cite{monotone}. On the other hand, since most applications of
quantum information theory require the maximally entangled state for
faithfully transmission of quantum data, it is necessary to develop the
special technique of entanglement manipulation which uses local quantum
operations and classical communication to purify impure entangled states 
\cite{procus,bennett2}.

There has been extensive work on entanglement manipulation. For the case of
two-qubit systems, the Procrustean method \cite{procus} provides local
operations to obtain the maximally entangled state from a partly entangled
pure state. The general theory of entanglement transformations for pure
states of the bipartite system has also been proposed \cite{transfor}-\cite
{transfor2}. As far as tripartite states \cite{tripart}-\cite{tripart1} are
concerned, researches have shown that the GHZ state \cite{ghz} is the
maximally entangled state which violates Bell inequalities maximally and
maximize the mutual information of local measurements \cite{proof}. Hence,
it is desirable to propose a general algorithm to carry out the local
operations to distill the GHZ state from a given tripartite state. In this
paper we present such an algorithm. Using this algorithm, we obtain an
analytical result of the distillation operations for the general state of
three qubits. As far as we know, it is the first time for the result of this
type to be achieved.

What we address is to transform a tripartite state $|\psi _{ABC}\rangle $ of
three qubits A, B and C into the state $|\psi _{GHZ}\rangle \equiv \frac 1{%
\sqrt{2}}(|000\rangle +|111\rangle )$ by local actions 
\begin{equation}
\frac 1{N^{1/2}}T_A\otimes T_B\otimes T_C|\psi _{ABC}\rangle =|\psi
_{GHZ}\rangle ,  \label{summ}
\end{equation}
where $N=\langle \psi _{ABC}|T_A^{\dagger }T_A\otimes T_B^{\dagger
}T_B\otimes T_C^{\dagger }T_C|\psi _{ABC}\rangle $ is the probability that
the manipulation succeeded. Once the operators $T_A$, $T_B$ and $T_C$ are
worked out, one can construct the simplest local operations---generalized
measurements or actions of local filters on qubits A, B, and C---to
implement the distillation process optimally.

We first give a special representation for the three-qubit state $|\psi
_{ABC}\rangle $, which will be useful in our distillation method. The
representation, which we call ``Wootters' representation'' here, is
introduced according to the following idea. Suppose, the tripartite state $%
|\psi _{ABC}\rangle $ considered has a form 
\begin{equation}
|\psi _{ABC}\rangle =\sum_{i=0}^1|i_A\rangle |\varphi _i^{\bar{A}}\rangle ,
\label{decom}
\end{equation}
where $\{|0_A\rangle ,|1_A\rangle \}$ are the standard basis of qubit A, and 
$|\varphi _i^{\bar{A}}\rangle $s are states of the ensemble which realizes
the reduced density matrix $\rho _{BC}$ of qubits B and C. Note that $%
|\varphi _i^{\bar{A}}\rangle $ is subnormalized, namely, the squared length $%
\langle \varphi _i^{\bar{A}}|\varphi _i^{\bar{A}}\rangle $ is equal to the
probability of $|\varphi _i^{\bar{A}}\rangle $ in the ensemble. We then can
get a kind of representations of $|\psi _{ABC}\rangle $ by different choices
of the basis of qubit A 
\begin{equation}
|\psi _{ABC}\rangle =\sum_{i=0}^1|i_A^{\prime }\rangle |\phi _i^{\bar{A}%
}\rangle  \label{decom2}
\end{equation}
with 
\begin{equation}
|i_A^{\prime }\rangle =\hat{V}_A^{\intercal }|i_A\rangle
=\sum_{j=0}^1(V_A)_{ij}|j_A\rangle ,\ |\phi _i^{\bar{A}}\rangle
=\sum_{j=0}^1(V_A^{*})_{ij}|\varphi _j^{\bar{A}}\rangle .  \label{decorel}
\end{equation}
Here $i=0,1$ and the transposition is taken in the standard basis $%
\{|i_A\rangle \}$ (it is included in order to be consistent with previous
papers, e.g., see Ref. \cite{wootters1}). The matrix $V_A$ is a
representation of the unitary operator $\hat{V}_A$ in the standard basis.
Introduce the time reversal operation for two-qubit states with $|\tilde{\phi%
}\rangle =\sigma _2\otimes \sigma _2|\phi ^{*}\rangle $, and define the
symmetric matrix $\tau _{\phi ^{\bar{A}}}$ for $\{|\phi _i^{\bar{A}}\rangle
\}$ with 
\begin{equation}
\tau _{\phi ^{\bar{A}}}^{ij}=\langle \phi _i^{\bar{A}}|\tilde{\phi}_j^{\bar{A%
}}\rangle ,\ \ \ i=0,1.  \label{tdef}
\end{equation}
Then according to the studies of Wootters \cite{wootters1}, there exists a
representation 
\begin{equation}
|\psi _{ABC}\rangle =\sum_{i=0}^1|A_i\rangle |x_i^{\bar{A}}\rangle
\label{repres1}
\end{equation}
with 
\begin{equation}
|A_i\rangle =\hat{U}_A^{\intercal }|i_A\rangle
=\sum_{j=0}^1(U_A)_{ij}|j_A\rangle ,\ |x_i^{\bar{A}}\rangle
=\sum_{j=0}^1(U_A^{*})_{ij}|\varphi _j^{\bar{A}}\rangle  \label{decorel2}
\end{equation}
such that 
\begin{equation}
\tau _{x^{\bar{A}}}^{ij}=(U_A\tau _{\varphi ^{\bar{A}}}U_A^{\intercal
})_{ij}=\delta _{ij}\pi _i^{\bar{A}},\ \ \ i,j=0,1.  \label{trelat}
\end{equation}
Here the parameters $\pi _i^{\bar{A}}$s, which satisfy $\pi _0^{\bar{A}}\geq
\pi _1^{\bar{A}}\geq 0$, are the square roots of the eigenvalues of the
Hermitian matrix $\tau _{\phi ^{\bar{A}}}\tau _{\phi ^{\bar{A}}}^{*}$ \cite
{book}. The difference of them defines a ``concurrence''\cite
{wootters1,wootters2} which provides a measure of entanglement for the
two-qubit state $\rho _{BC}$.

Similarly, we can write down two other Wootters' representations of the
state $|\psi _{ABC}\rangle $ 
\[
|\psi _{ABC}\rangle =\sum_{i=0}^1|i_B\rangle |\varphi _i^{\bar{B}}\rangle
=\sum_{i=0}^1|B_i\rangle |x_i^{\bar{B}}\rangle , 
\]
\begin{equation}
|\psi _{ABC}\rangle =\sum_{i=0}^1|i_C\rangle |\varphi _i^{\bar{C}}\rangle
=\sum_{i=0}^1|C_i\rangle |x_i^{\bar{C}}\rangle  \label{repres2}
\end{equation}
with relations 
\[
|R_i\rangle =\hat{U}_R^{\intercal }|i_R\rangle
=\sum_{j=0}^1(U_R)_{ij}|j_R\rangle ,\ |x_i^{\bar{R}}\rangle
=\sum_{j=0}^1(U_R^{*})_{ij}|\varphi _j^{\bar{R}}\rangle , 
\]
\begin{equation}
\langle x_i^{\bar{R}}|\tilde{x}_j^{\bar{R}}\rangle =\delta _{ij}\pi _i^{\bar{%
R}},\ \ \ \ \ R=B,C,\ \ \ i=0,1.  \label{rel3}
\end{equation}

Our distillation algorithm makes use of a set of local operations 
\[
f_R=(\frac{\pi _1^{\bar{R}}}{\pi _0^{\bar{R}}})^{1/2}|R_0\rangle \langle
R_0|+|R_1\rangle \langle R_1| 
\]
\begin{equation}
=U_R^{\intercal }[(\frac{\pi _1^{\bar{R}}}{\pi _0^{\bar{R}}}%
)^{1/2}|0_R\rangle \langle 0_R|+|1_R\rangle \langle 1_R|]U_R^{*},\ R=A,B,C.
\label{operation}
\end{equation}
Note that these operations require $\pi _1^{\bar{R}}\neq 0$ since for the
case of $\pi _1^{\bar{R}}=0$, $f_R$ becomes a projection operator of the
pure state of qubit R and it shall disentangle the tripartite state. In
fact, the states with $\pi _1^{\bar{R}}=0$ cannot be distilled to the GHZ
state by any local actions and classical communication acting individually
on them. Further discussions shall be presented later in this paper.

Direct observation can be found that the action $f_R$ on the qubit $R$ ($%
R=A,B,C$) causes the reduced matrix of the two other qubits to be separable.
For example, consider the operation $f_A$ on the qubit A. According to the
Wootters' representation (\ref{repres1}), after the operation, one shall
acquire a state (note that such a state is obtained probalistically) with $%
\pi _0^{\bar{A}}=\pi _1^{\bar{A}}$. This means that the reduced matrix $\rho
_{BC}$ has a zero concurrence, thus is nonentangled.

Theorem 1: A tripartite entangled state of three qubits $|\psi _{ABC}\rangle 
$ which has Wootters' representations with $\pi _1^{\bar{R}}\neq 0$ ($%
R=A,B,C $) can be transformed to the generalized GHZ state by the local
operations 
\begin{equation}
\ f_A\otimes f_B\otimes f_C|\psi _{ABC}\rangle ,  \label{trans}
\end{equation}
where the operators $f_A$, $f_B$, and $f_C$ are given by Eq. (\ref{operation}%
). Note that a generalized GHZ state $|\psi _{GGHZ}\rangle $ has the
following form 
\begin{equation}
|\psi _{GGHZ}\rangle =\alpha |000\rangle +\beta |111\rangle  \label{gghz}
\end{equation}
with the parameters $0\leq \alpha \leq \beta $ by a suitable choice of the
local basis.

Proof: Consider the reduced density matrix $\rho _{BC}$ of qubits B and C.
After the local operation $f_A$, it becomes a separable state. Since the
local operations $f_B$ and $f_C$ can not produce any entanglement for the
nonentangled state of qubits B and C, it shall still be a separable state in
the final outcome of the set of actions (\ref{trans}). According to the
communicative property of the set of operators $\{f_A,f_B,f_C\}$, the same
analysis is applicable to the reduced density of qubit pairs A-B and A-C.
Thus, after the local operations (\ref{trans}), we arrive at a state which
is pairwise separable. We now need to show that a pairwise separable state
of three qubits is a generalized GHZ state.

Lemma 1: A tripartite entangled state of three qubits is pairwise separable
if and only if it is a generalized GHZ state.

Proof: It is obvious that a generalized GHZ state (\ref{gghz}) is pairwise
separable. To prove the converse statement, we use the fact that an
arbitrary tripartite state of three qubits can be written in the form \cite
{minimum} 
\begin{equation}
|\psi _{ABC}\rangle =\lambda _0|000\rangle +\lambda _1e^{i\varphi
}|100\rangle +\lambda _2|101\rangle +\lambda _3|110\rangle +\lambda
_4|111\rangle  \label{minimum}
\end{equation}
by a suitable choice of local basis, where the coefficients $\lambda _i$ are
all real and non-negative and $\varphi $ is a phase between $0$ and $\pi $.
This representation is a minimal description, with only five terms, for the
three-qubit states. Now, suppose a tripartite entangled state $|\psi
_{ABC}^s\rangle $ which has a minimal representation with coefficients $%
\lambda _i^s$ $(i=0,\cdots ,4)$ and a phase $\varphi ^s$ is pairwise
separable. Then, due to separability, there are relations $\pi _0^{\bar{R}%
}=\pi _1^{\bar{R}}\ (R=A,B,C)$ for the Wootters' representations of $|\psi
_{ABC}^s\rangle $. A straightforward calculation yields 
\begin{equation}
\lambda _2^s\lambda _3^s-\lambda _1^s\lambda _4^se^{i\varphi ^s}=0,
\label{condips0}
\end{equation}
\begin{equation}
\lambda _0^s\lambda _2^s=0,  \label{condips1}
\end{equation}
\begin{equation}
\lambda _0^s\lambda _3^s=0.  \label{condips}
\end{equation}
Noticing that $|\psi _{ABC}^s\rangle $ is a tripartite entangled state,
i.e., it can not be written as a biseparable form and a product form of the
three qubits, we then have the solutions $\lambda _1^s=\lambda _2^s=\lambda
_3^s=0$ for the above set of equations. Now the state $|\psi _{ABC}^s\rangle 
$ takes the form 
\begin{equation}
|\psi _{ABC}^s\rangle =\lambda _0^s|000\rangle +\lambda _4^s|111\rangle .
\label{gghz2}
\end{equation}
This completes our proof of Lemma 1 and then Theorem 1.

Let $|\psi _{ABC}^{\prime }\rangle $ denote the state of the outcome
corresponding to the local operations of (\ref{trans}). According to Theorem
1, we have 
\begin{equation}
|\psi _{ABC}^{\prime }\rangle =\alpha |0^{\prime }0^{\prime }0^{\prime
}\rangle +\beta |1^{\prime }1^{\prime }1^{\prime }\rangle .  \label{gghz3}
\end{equation}
The coefficients $\alpha $, $\beta $ and the local basis $\{|0^{\prime
}\rangle ,|1^{\prime }\rangle \}$ of the three qubits are completely
determined by the initially given state $|\psi _{ABC}\rangle $. They can be
directly calculated from Eqs. (\ref{operation}) and (\ref{trans}). Now to
acquire the GHZ state one only needs to carry out one of the following
actions (noticing that we have set $\alpha \leq \beta $) 
\begin{equation}
f_R^{\prime }=|0_R^{\prime }\rangle \langle 0_R^{\prime }|+\frac \alpha \beta
|1_R^{\prime }\rangle \langle 1_R^{\prime }|,\ R=A,B,C  \label{ggope}
\end{equation}
on the qubits A, B or C respectively.

The algorithm established above shall enable one to achieve local operators
of the distillation actions for any pure tripartite state of three qubits.
Before presenting the general result for such distillation actions, we give
some discussions which our algorithm implies.

In general, to distill the GHZ state from a three-qubit state requires local
operations on each of the three qubits. Nevertheless, there are exceptions.
The simplest example is the generalized GHZ state for which one only needs
to perform a local operation on any one of the three qubits. In detail, on
which qubits the operations to be performed is determined by properties of
the three reduced matrices $\rho _{AB}$, $\rho _{AC}$, and $\rho _{BC}$%
---whether they are separable or not. For instance, assume $\rho _{AB}$ of
the state $|\psi _{ABC}\rangle $ is separable. Then we have $\pi _0^{\bar{C}%
}=\pi _1^{\bar{C}}$ due to its separability. This results in that the local
operator $f_C$ of Eq. (\ref{operation}) becomes an identity operator. A
noticeable example of this type is the slice state \cite{slice} which has
the form 
\begin{equation}
|\psi _{slice}\rangle =\lambda _0|000\rangle +\lambda _1|100\rangle +\lambda
_4|111\rangle .  \label{slice}
\end{equation}
A simple calculation shows that the reduced matrices $\rho _{AB}$ and $\rho
_{AC}$ of it are separable. Thus to achieve the GHZ state from $|\psi
_{slice}\rangle $ one only needs to perform a local operation $T_A$ on the
qubit A.

Now we present an analytical result of distillation operators $T_A\otimes
T_B\otimes T_C$ in (\ref{summ}) for the general pure three-qubit state. We
still use the minimal representation (\ref{minimum}) for the state $|\psi
_{ABC}\rangle $. Direct calculations give the parameters of the state $|\psi
_{ABC}\rangle $ 
\[
\pi _{0,1}^{\bar{A}}=\sqrt{|\Delta |^2+\lambda _0^2\lambda _4^2}\pm |\Delta
|, 
\]
\[
\pi _{0,1}^{\bar{B}}=\lambda _0(\sqrt{\lambda _2^2+\lambda _4^2}\pm \lambda
_2), 
\]
\begin{equation}
\pi _{0,1}^{\bar{C}}=\lambda _0(\sqrt{\lambda _3^2+\lambda _4^2}\pm \lambda
_3),  \label{parameter}
\end{equation}
where `$+$' is for $\pi _0^{\bar{R}}$ and `$-$' for $\pi _1^{\bar{R}}$. The
complex number $\Delta $ is defined as 
\begin{equation}
\Delta \equiv \lambda _2\lambda _3-\lambda _1\lambda _4e^{i\varphi }.
\label{deta}
\end{equation}
The local operations which transform the state $|\psi _{ABC}\rangle $ into
the GHZ state are given by 
\begin{equation}
T_A\otimes T_B\otimes T_C=U_A^{\prime }f_A\otimes U_B^{\prime }f_B\otimes
U_C^{\prime }f_C^{\prime }f_C,  \label{operator}
\end{equation}
where the operators are shown as follows:

$f_R$: $f_R$ ($R=A,B,C$) is given by Eq. (\ref{operation}). The unitary
operator $\hat{U}_R$ in it comprises two parts 
\begin{equation}
U_R=U_{R1}U_{R0}.  \label{preunitar}
\end{equation}
Here, $\hat{U}_{R0}$ is a unitary transformation which diagonalizes the
Hermitian matrix $\tau _{\varphi ^{\bar{R}}}\tau _{\varphi ^{\bar{R}}}^{*}$.
Direct calculations give that 
\begin{equation}
U_{A0}=e^{-i\theta _2\sigma _2}e^{-i\theta _3\sigma _3},U_{B0}=e^{-i\theta
_B\sigma _2},U_{C0}=e^{-i\theta _C\sigma _2},  \label{ur0}
\end{equation}
where 
\[
\theta _2=\arctan \frac{|\Delta |-\sqrt{|\Delta |^2+\lambda _0^2\lambda _4^2}%
}{\lambda _0\lambda _4},\ \theta _3=-\arctan \frac{|\Delta |+%
%TCIMACRO{\func{Re} }
%BeginExpansion
\mathop{\rm Re}%
%EndExpansion
\Delta }{%
%TCIMACRO{\func{Im} }
%BeginExpansion
\mathop{\rm Im}%
%EndExpansion
\Delta }, 
\]
\begin{equation}
\theta _B=\arctan \frac{\lambda _2-\sqrt{\lambda _2^2+\lambda _4^2}}{\lambda
_4},\ \theta _C=\arctan \frac{\lambda _3-\sqrt{\lambda _3^2+\lambda _4^2}}{%
\lambda _4}.  \label{thetas}
\end{equation}
The operator 
\begin{equation}
U_{R1}=i|1_R\rangle \langle 1_R|+|0_R\rangle \langle 0_R|  \label{ur1}
\end{equation}
is included so that the diagonal elements of $U_R\tau _{\varphi ^{\bar{R}%
}}U_R^{\intercal }$ become real and non-negative.

$f_R^{\prime }$: The operator $f_R^{\prime }$ ($R=A$, $B$, or $C$) of (\ref
{ggope}) transforms the generalized GHZ state into the GHZ state. It is
equivalent to perform any one of them. In equation (\ref{operator}) we have
chosen the operation $f_C^{\prime }$ acting on the qubit C. To achieve $%
f_R^{\prime }$ requires the detailed knowledge of parameters and local basis
of the generalized GHZ state (\ref{gghz3}). After some hard but
straightforward calculations, we obtain the ratio of the two coefficients 
\begin{equation}
\frac \alpha \beta =\left[ \frac{(\lambda _2^2+\lambda _4^2)(\lambda
_3^2+\lambda _4^2)}{|\Delta |^2+\lambda _0^2\lambda _4^2}\right] ^{1/2}
\label{ratio}
\end{equation}
and the local basis of the qubit C 
\[
|0_C^{\prime }\rangle =\frac{\sqrt{2}i}2[(\cos \theta _C+\sin \theta
_C)|0_C\rangle +(\cos \theta _C-\sin \theta _C)|1_C\rangle ], 
\]
\begin{equation}
\ |1_C^{^{\prime }}\rangle =\frac{\sqrt{2}i}2[(\cos \theta _C-\sin \theta
_C)|0_C\rangle -(\cos \theta _C+\sin \theta _C)|1_C\rangle ].  \label{localc}
\end{equation}
Here we assume $(\lambda _2^2+\lambda _4^2)(\lambda _3^2+\lambda _4^2)\leq
|\Delta |^2+\lambda _0^2\lambda _4^2$, and the parameter $\Delta $ is given
in Eq. (\ref{deta}) and $\theta _C$ in (\ref{thetas}).

$U_R^{\prime }$: The final local unitary transformations 
\begin{equation}
U_R^{\prime }=|0_R\rangle \langle 0_R^{\prime }|+|1_R\rangle \langle
1_R^{\prime }|\ \ \ R=A,B,C  \label{finunitar}
\end{equation}
are included to revert the local basis $\{|0_R^{\prime }\rangle
,|1_R^{\prime }\rangle \}$ of generalized GHZ state (\ref{gghz3}) to the
initial standard basis $\{|0_R\rangle ,|1_R\rangle \}$ of the state $|\psi
_{ABC}\rangle $. $\{|0_C^{\prime }\rangle ,|1_C^{\prime }\rangle \}$ have
been given in Eq. (\ref{localc}). Now we present the local basis of qubits A
and B 
\[
|0_A^{\prime }\rangle =\frac{\sqrt{2}i}2[(\cos \theta _2+\sin \theta
_2)e^{-i\theta _3}|0_A\rangle +(\cos \theta _2-\sin \theta _2)e^{i\theta
_3}|1_A\rangle ], 
\]
\[
|1_A^{\prime }\rangle =\frac{\sqrt{2}i}2[(\cos \theta _2-\sin \theta
_2)e^{-i\theta _3}|0_A\rangle -(\cos \theta _2+\sin \theta _2)e^{i\theta
_3}|1_A\rangle ], 
\]
\[
|0_B^{\prime }\rangle =\frac{\sqrt{2}i}2[(\cos \theta _B+\sin \theta
_B)|0_B\rangle +(\cos \theta _B-\sin \theta _B)|1_B\rangle ], 
\]
\begin{equation}
\ |1_B^{^{\prime }}\rangle =\frac{\sqrt{2}i}2[(\cos \theta _B-\sin \theta
_B)|0_B\rangle -(\cos \theta _B+\sin \theta _B)|1_B\rangle ],
\label{localab}
\end{equation}
where $\theta _2$, $\theta _3$, and $\theta _B$ are given by (\ref{thetas}).

There remains one case to consider, namely, the states with parameters $\pi
_1^{\bar{R}}=0$ $(R=A,B,C)$ for their Wootters' representations. For these
states our distillation algorithm is not valid any more. In fact, as far as
the tripartite entangled state is concerned, the three relations $\pi _1^{%
\bar{R}}=0$ for $R=A,B$ and $C$ are equivalent. For example, the condition $%
\pi _1^{\bar{A}}=0$ implies that $\pi _1^{\bar{B}}=\pi _1^{\bar{C}}=0$. This
can be seen from Eq. (\ref{parameter}). Assume $\pi _1^{\bar{A}}=0$ for the
state $|\psi _{ABC}\rangle $. Since $\lambda _0\neq 0$ (otherwise the state $%
|\psi _{ABC}\rangle $ will be biseparable), the relation $\pi _1^{\bar{A}}=0$
leads to $\lambda _4=0$, and then $\pi _1^{\bar{B}}=\pi _1^{\bar{C}}=0$. The
state $|\psi _{ABC}\rangle $ now has the form 
\begin{equation}
|\psi _{ABC}\rangle =\lambda _0|000\rangle +\lambda _1e^{i\varphi
}|100\rangle +\lambda _2|101\rangle +\lambda _3|110\rangle .  \label{wclass}
\end{equation}
This state is the so-called ``W-class state'' \cite{wclass} which could not
be distilled to the GHZ state by any local actions. It can be understood in
the following way. Since, the ratio $\pi _1^{\bar{A}}/\pi _0^{\bar{A}}$ is a
constant under the invertible local operations of qubits B and C (see Ref. 
\cite{ratio,ratio1}), the relation $\pi _1^{\bar{A}}=\pi _1^{\bar{B}}=\pi
_1^{\bar{C}}=0$ shall be retained under the actions $T_B\otimes T_C$.
Similar analysis gives that this relation shall also hold under the action
of $T_A$. Thus we can conclude that local operations $T_A\otimes T_B\otimes
T_C$ shall take a W-class state to another W-class state, so that the GHZ
state will never be produced.

In summary, we have presented an algorithm to distill the GHZ state from a
single copy of the three-qubit state. It enable one to achieve directly the
operators of the distillation operations, thus the distillation process can
be realized optimally. We then apply our distillation algorithm to the
general state of three qubits and obtain an analytical result of operations
for such a process. Finally, we show that the state for which our
distillation algorithm is not valid cannot be distilled to the GHZ state by
any local operations. Thus the distillation algorithm we presented is
sufficient for the tripartite state of three qubits.

\begin{center}
{\bf Acknowledgments}
\end{center}

This work was supported in part by the National Natural Science Foundation,
the Doctoral Education Fund of the Education Ministry, and the Nuclear
Theory Fund of HIRFL of China.


\begin{thebibliography}{99}
\bibitem{bennett1}  C.H. Bennett, Phys. Today {\bf 48}, No. 10, 24 (1995).

\bibitem{procus}  C.H. Bennett, H. Bernstein, S. Popescu, and B. Schumacher,
Phys. Rev. A {\bf 53}, 2046 (1996).

\bibitem{monotone}  G. Vidal, Journ. of Mod. Opt. {\bf 47}, 355 (2000).

\bibitem{bennett2}  C. H. Bennett, G. Brassard, S. Popescu, B. Schumacher,
J.A. Smolin, and W.K. Wootters, Phys. Rev. Lett. {\bf 76}, 722 (1996).

\bibitem{transfor}  M.A. Nielsen, Phys. Rev. Lett. {\bf 83}, 436 (1999).

\bibitem{transfor1}  G. Vidal, Phys. Rev. Lett. {\bf 83}, 1046 (1999).

\bibitem{transfor2}  D. Jonathan and M.B. Plenio, Phys. Rev. Lett. {\bf 83},
1455 (1999).

\bibitem{tripart}  O. Cohen and T.A. Brun, Phys. Rev. Lett. {\bf 84}, 5908
(2000).

\bibitem{minimum}  A. Ac\'{i}n, A. Andrianov, L. Costa, E. Jan\'{e}, J.I.
Latorre, and R. Tarrach, Phys. Rev. Lett. {\bf 85}, 1560 (2000).

\bibitem{wclass}  W. D\"{u}r, G. Vidal and J.I. Cirac, quant-ph/0005115.

\bibitem{tripart1}  T.A. Brun and O. Cohen, quant-ph/0005124.

\bibitem{ghz}  D.M. Greengerger, M. Horne, A. Zeilinger, {\it Bell's
theorem, Quantum Theory, and the Conceptions of the Universe}, ed. M.
Kafatos, Kluwer, Dordrecht 69 (1989).

\bibitem{proof}  N. Gisin and H. Beschmann-Pasquinucci, Phys. Lett. A {\bf %
246}, 1 (1998).

\bibitem{wootters1}  W.K. Wootters, Phys. Rev. Lett. {\bf 80}, 2245 (1998).

\bibitem{book}  R.A. Horn and C.R. Johnson, {\it Matrix Analysis} (Cambridge
University Press, New York, 1985), p. 205.

\bibitem{wootters2}  S. Hill and W.K. Wootters, Phys. Rev. Lett. {\bf 78},
5022 (1997).

\bibitem{slice}  H.A. Carteret and A. Sudbery, J. Phys. A {\bf 33}, 4981
(2000).

\bibitem{ratio}  N. Linden, S. Massar, and S. Popescu, Phys. Rev. Lett. {\bf %
81}, 3279 (1998).

\bibitem{ratio1}  A. Kent, N. Linden, and S. Massar, Phys. Rev. Lett. {\bf 83%
}, 2656 (1999).
\end{thebibliography}
\end{document}